\documentclass[sn-mathphys-num]{sn-jnl}% Math and Physical Sciences Numbered Reference Style 
\usepackage{graphicx}%
\usepackage{multirow}%
\usepackage{amsmath,amssymb,amsfonts}%
\usepackage{amsthm}%
\usepackage{mathrsfs}%
\usepackage[title]{appendix}%
\usepackage{xcolor}%
\usepackage{textcomp}%
\usepackage{manyfoot}%
\usepackage{booktabs}%
\usepackage{algorithm}%
\usepackage{algorithmicx}%
\usepackage{algpseudocode}%
\usepackage{listings}%
\usepackage{adjustbox}
\theoremstyle{thmstyleone}%
\newtheorem{theorem}{Theorem}%  meant for continuous numbers
\newtheorem{proposition}[theorem]{Proposition}% 

\newtheorem{lemma}[theorem]{Lemma}

\theoremstyle{thmstyletwo}%
\newtheorem{example}{Example}%

\theoremstyle{thmstylethree}%
\newtheorem{definition}{Definition}%

\begin{document}

\title[Construction of Additive Complementary Dual Codes Over Finite Fields]{Construction of Additive Complementary Dual Codes Over Finite Fields}


\author[1]{\fnm{Gyanendra K.} \sur{Verma} \footnote{ Orcid ID: Gyanendra K. Verma - \url{https://orcid.org/0000-0001-5872-7702}}}\email{gkvermaiitdmaths@gmail.com}

\author[2]{\fnm{R. K.} \sur{Sharma}\footnote{Orcid ID: R. K. Sharma - \url{https://orcid.org/0000-0001-5666-4103}}}\email{rksharmaiitd@gmail.com}


\affil*[1]{\orgdiv{Department of Mathematics}, \orgname{Indian Institute of Technology Delhi}, \orgaddress{\street{Huaz Khas}, \city{New Delhi}, \postcode{110016}, \state{Delhi}, \country{India}}}

\affil[2]{\orgdiv{Department of Mathematics}, \orgname{South Asian University}, \orgaddress{\street{} \city{New Delhi}, \postcode{110068}, \state{Delhi}, \country{India}}}
%
%\affil[3]{\orgdiv{Department}, \orgname{Organization}, \orgaddress{\street{Street}, \city{City}, \postcode{610101}, \state{State}, \country{Country}}}


\abstract{In this work, we investigate additive complementary dual (ACD) codes and their construction over finite fields $\mathbb{F}_{q^2}$ with respect to the trace inner products, where $q$ is a prime power. First, we associate an additive code with a matrix known as a generator matrix. After that, we describe ACD codes in terms of generator matrices for the trace Hermitian and the trace Euclidean inner products. We also construct ACD codes over $\mathbb{F}_{q^2}$ from linear codes over $\mathbb{F}_q.$ Additionally, we present techniques for constructing ACD codes with various parameters from a given ACD code over $\mathbb{F}_{q^2}.$ By applying these methods, we construct numbers of trace Euclidean and trace Hermitian ACD codes that exhibit better parameters compared to the best known linear codes over $\mathbb{F}_9$ and $\mathbb{F}_4$ of the same size and length.}




\maketitle
\section{Introduction}\label{sec1}
Coding theory investigates the construction and analysis of error-correcting codes which are essential for reliable data transmission through noisy channels. Linear codes are an important class of code in this line of work due to their algebraic structure and error-correcting abilities. Additive codes are a versatile extension of linear codes that provide more flexibility and efficiency in error correction, making them useful in a wide range of communication and information theory applications. Due to the desirable algebraic properties of finite fields and finite rings, many scholars have focused on the study of linear code over these alphabets. In particular, researchers extensively investigated linear complementary dual (LCD) codes over finite fields due to their wide range of applications in consumer electronics, data storage, and communication systems. LCD codes can be employed to counter side-channel and fault injection attacks, as highlighted in \cite{carlet2016}. As discussed in \cite{masseylcd}, LCD codes offer the optimal linear coding solution for the binary adder channel with two users. These codes were defined and characterized using generator matrices by Massey \cite{masseyreversible}.

The concept of additive codes in terms of association schemes was first proposed by Delsarte \cite{fadditive,additivebook} in 1973. Borges et al. \cite{z2z4}
defined $\mathbb{Z}_2\mathbb{Z}_4$-additive codes as $\mathbb{Z}_4$-submodules of $\mathbb{Z}_2^{\alpha}\times \mathbb{Z}_4^{\beta}.$  In 2021, Shi et al. \cite{shi2021}  decomposed $\mathbb{Z}_2\mathbb{Z}_4$-additive quasi-cyclic codes   into constituent codes and derived criteria for self-duality and linear complementarity. Aydogdu and Siap \cite{z2z2sac} extended  $\mathbb{Z}_2\mathbb{Z}_4$-additive codes to $\mathbb{Z}_2 \mathbb{Z}_{2^s}$-additive codes. For these codes, they also provided generator and parity check matrices. Aydogdu et al. \cite{z2z2u} introduced $\mathbb{Z}_2\mathbb{Z}_2[u]$-additive codes and established the standard form of generator  and parity check matrices for these codes. Bandi et al. \cite{bandi2017} described the structure of negacyclic codes and their dual codes of length $2^k$ over $\mathbb{Z}_4+u\mathbb{Z}_4.$ They also used Gray maps to come up with good $\mathbb{Z}_4$-linear codes. In 2018, Shi et al. \cite{shi2018} investigated $\mathbb{Z}_p\mathbb{Z}_{p^k} $-Additive codes using Gray-like maps and constructed $1$-perfect additive code in mixed alphabets. Verma et al. \cite{omadd} studied skew-additive constacyclic codes over a certain type of non-chain rings and constructed a number of quantum codes.
Debnath et al. \cite{omadd2} investigated additive constacyclic codes over Frobenius rings and constructed quantum codes by applying CSS construction. Shi et al. \cite{lcdacdunital} constructed LCD and ACD codes over a non-commutative non-unital ring and showed that, as a special case, ACD codes include LCD codes. Linear codes are additive codes; therefore, LCD codes are ACD codes. In 2022, Shi et al. \cite{acf4} studied additive cyclic complimentary dual codes over the finite field $\mathbb{F}_4.$ The authors derived the generator polynomials for the dual of additive cyclic codes and also examined the trace and subfield codes of these codes. In the same year, Shi et al. \cite{af4} explored ACD codes over $\mathbb{F}_4.$ They described ACD codes in terms of generator matrices and constructed ACD codes over $\mathbb{F}_4$ from binary codes with respect to the trace Euclidean and the trace Hermitian inner products.

For security purposes, using ACD codes is a viable option. The self-orthogonal additive codes with respect to the trace inner products and symplectic inner product are used to construct good quantum error-correcting codes \cite{ketkar2006,calderbank1998}. Due to the significant role of additive codes  in both security applications and quantum codes, in this article, we investigate additive complementary dual (ACD) codes over finite fields $\mathbb{F}_{q^2},$ where $q$ is a prime power. Moreover, we provide methods for constructing ACD codes over $\mathbb{F}_{q^2}.$ Using these methods, we construct several trace Euclidean and trace Hermitian ACD codes over $\mathbb{F}_9$ and $\mathbb{F}_4.$ These codes possess better parameters than the best-known linear codes of the same size and length. We organize the remainder of the paper as follows.

Section \ref{pre} of this paper covers the basics of additive codes with respect to two trace inner products and the association of a matrix with an additive code. Section \ref{trhacd} focuses on trace Hermitian ACD codes, where we provide a characterization of these codes in terms of generator matrices over finite fields and present a construction of ACD codes over $\mathbb{F}_{q^2}$. In Section \ref{treacd}, we characterize and construct trace Euclidean ACD codes over $\mathbb{F}_{q^2}$ from linear codes. In Section \ref{construction acd}, we present methods for constructing several ACD codes with parameters $(n,q^{k+1})_{q^2}$ and $(n+1,q^{k+1})_{q^2}$ using ACD code with parameters $(n,q^k)_{q^2}$ over $\mathbb{F}_{q^2}$. In Section \ref{conclusion}, we conclude the manuscript. 

\section{Preliminaries}\label{pre}
In this paper, we denote $\mathbb{F}_{q^2}$  as a finite field with $q^2$  elements, where $q$ is of the form $p^e$, with $p$ being a prime and $e$ being a positive integer. For an element $b \in \mathbb{F}_{q^2}$ , we define its conjugate as $\overline{b}=b^q$. Moreover, for an element $b=(b_1,b_2,\dots,b_n)\in \mathbb{F}_{q^2}^n$, we denote its conjugate as $\overline{b}=(\overline{b_1},\overline{b_2},\dots,\overline{b_n}).$ The \textit{trace map} $Tr: \mathbb{F}_{q^2} \to \mathbb{F}_{q}$ is defined by 
$Tr(b)=b+b^q\ \  \forall b\in \mathbb{F}_{q^2} .$ Note that the \textit{trace map} is surjective over $\mathbb{F}_{q}$ and satisfies $Tr(b+c)=Tr(b)+Tr(c)$ for all $b,c\in \mathbb{F}_{q^2}. $  The \textit{Hamming weight} of an element $x=(x_1,x_2,\dots,x_n)\in \mathbb{F}_{q^2}^n$  is denoted by $wt(x)$ and  defined as the number of nonzero coordinates in $x.$  The \textit{Hamming distance} between any two elements $x=(x_1,x_2,\dots,x_n)$ and $y=(y_1,y_2,\dots,y_n)$ in $\mathbb{F}_{q^2}^n$ is defined as the Hamming weight of $x-y.$

\begin{definition}
	An additive subgroup of $\mathbb{F}_{q^2}^n$ is said to be an additive code over $\mathbb{F}_{q^2}$ of length $n.$
\end{definition}
\begin{definition}
	The minimum distance of code $C$ is the smallest number of coordinates at which any two distinct codewords in $C$ differ.
\end{definition}
We denote $(n, M,d)_{q^2}$, the parameters of an additive code over $\mathbb{F}_{q^2},$ where $n$ is the length of the codewords, $M$ is the size of $C$, and $d$ is the minimum distance of $C$.  We write the parameters of a linear code $C$ as $[n,k,d]_{q^2}$  if $C$ is a $k$-dimensional subspace of $\mathbb{F}_{q^2}^n$ with the minimum distance $d.$
An additive code over $\mathbb{F}_{q^2}$  always has a basis over the base field $\mathbb{F}_p.$ If $C$ is an additive $(n,q^{k})_{q^2}$ code, then its basis contains $ek$ codewords over $\mathbb{F}_p$. Henceforth, we denote $t=ek$ for additive codes  over $\mathbb{F}_{q^2}$ with parameters $(n,q^k)_{q^2}.$

\begin{definition}
	A matrix $G$ of order $t\times n$ over $\mathbb{F}_{q^2}$  is called a generator matrix for an additive $(n,q^{k})$ code $C$ over $\mathbb{F}_{q^2}$ if $C=\{xG : x\in \mathbb{F}_p^{t}\}$.  The conjugate of a generator matrix $G=[g_{ij}]_{t\times n} $  is $\overline{G}=[\overline{g_{ij}}]_{t\times n}.$
\end{definition}

\begin{example}
	Let $\mathbb{F}_{9}=\{0,1,2,\beta,\beta +1,\beta +2,2\beta,2\beta +1,2\beta+2\},$ where $\beta^2+2\beta+2=0.$ Consider a linear $[5,3,1]_9$ code $C$ over $\mathbb{F}_{3^2}$ generated by the matrix 
	$$G=\begin{bmatrix}
		1 & 0 &\beta &0 &0\\
		0& 1&\beta &\beta &0\\
		0&0&0&0&\beta+1
	\end{bmatrix}.$$
	Then the code $C$ is also an additive $(5,3^6,1)_9$ code, and its generator  matrix corresponding to $G$ is
	$$\begin{bmatrix}
		1 & 0 &\beta &0 &0\\
		0&\beta &\beta+1&\beta+1&0\\
		0& 1&\beta &\beta &0\\
		0&0&0&0&\beta+1\\
		\beta &0&\beta+1&0&0\\
		0&0&0&0&2\beta+1
	\end{bmatrix}.$$
	Note that code $C$ has different  generator matrices of orders $3\times 5$ and $6\times 5$ in regard to linear and additive codes, respectively.
\end{example}
\begin{definition}
	Let $u=(u_1,u_2,\dots,u_n)$ and $w=(w_1,w_2,\dots,w_n)$ in $\mathbb{F}_{q^2}^n.$ Define the \textit{Euclidean inner product} and the \textit{trace Euclidean inner product} as 
	$$\langle u ,w\rangle_{E}=\sum_{j=1}^n u_jw_j \text{\ and\ } \langle u ,w\rangle_{TrE}=\sum_{j=1}^n Tr(u_jw_j),\text{\ respectively}.$$
\end{definition}
\begin{definition}
	Let $u=(u_1,u_2,\dots,u_n)$ and $w=(w_1,w_2,\dots,w_n)$ in $\mathbb{F}_{q^2}^n.$ The  \textit{Hermitian inner product} and the \textit{trace Hermitian inner product} are defined as 
	$$\langle u ,w\rangle_{H}=\sum_{j=1}^n u_j\overline{w_j}  \text{ and } \langle u ,w\rangle_{TrH}=\sum_{j=1}^n Tr(u_j\overline{w_j}), \text{ respectively}.$$
\end{definition}    

\begin{definition}
	Let $C$ be an $(n,q^k)_{q^2}$ additive code over $\mathbb{F}_{q^2}.$ The  \textit{Euclidean dual} $C^{\perp_{E}}$ is defined as $C^{\perp_{E}}=\{x\in \mathbb{F}_{q^2}^n \ : \langle x ,c\rangle_{E}=0\ \forall c\in C\}$  and the \textit{trace Euclidean dual} $C^{\perp_{TrE}}$ is defined as $C^{\perp_{TrE}}=\{x\in \mathbb{F}_{q^2}^n \ : \langle x ,c\rangle_{TrE}=0\ \forall c\in C\}.$ 
\end{definition}

\begin{definition}
	Let $C$ be an $(n,q^k)_{q^2}$ additive code over $\mathbb{F}_{q^2}.$ The \textit{Hermitian dual} $C^{\perp_{H}}$ is defined as $C^{\perp_{H}}=\{x\in \mathbb{F}_{q^2}^n \ : \langle x ,c\rangle_{H}=0\ \forall c\in C\}$ and the \textit{trace Hermitian dual} $C^{\perp_{TrH}}$ is defined as $C^{\perp_{TrH}}=\{x\in \mathbb{F}_{q^2}^n \ : \langle x ,c\rangle_{TrH}=0\ \forall c\in C\}.$ 
\end{definition}
\begin{definition}
	A linear code $C$ is called a Euclidean and a Hermitan linear  complementary dual (LCD) code if $C\cap C^{\perp_E}=\{0\}$ and  $C\cap C^{\perp_H}=\{0\},$ respectively.
\end{definition}
\begin{definition}
	An additive code $C$ is called a \textit{trace Euclidean additive complementary dual code} (ACD)  if $C\cap C^{\perp_{TrE}}=\{0\},$ and a \textit{trace Hermitian additive complementary dual code} (ACD) if $C\cap C^{\perp_{TrH}}=\{0\}.$
\end{definition}

Let $C^{\perp_{TrE}}$ be dual of an additive code defined with respect to the trace Euclidean  inner products. For $x,y\in C^{\perp_{TrE}}$,  $\langle x+y,c\rangle_{TrE}=\langle x,c\rangle_{TrE}+\langle y,c\rangle_{TrE}=0$ $\forall c\in C$, implies that $x+y\in C^{\perp_{TrE}}$. Hence, $C^{\perp_{TrE}}$ is also an additive code. Similarly, the trace Hermitian dual $C^{\perp_{TrH}}$ is an additive code. Moreover, if $C$ is an   $(n, q^k)$ over $\mathbb{F}_{q^2}$, then  $C^{\perp_{TrE}}$ and $C^{\perp_{TrH}}$ have parameters $(n, q^{2n-k})$.

The following result is an easy observation, indicating that for linear codes over $\mathbb{F}_{q^2},$ the trace Euclidean (and trace Hermitian) dual is equivalent to the Euclidean (and Hermitian) dual. Thus, studying linear codes with respect to trace inner products is redundant.

\begin{proposition}\label{hlcdtohacd}
	For a linear code $C$ over $\mathbb{F}_{q^2},$ the trace Hermitian dual $C^{\perp_{TrH}}$ is equal to the  Hermitian dual, and the trace Euclidean dual $C^{\perp_{TrE}}$ is equal to the Euclidean dual.
\end{proposition}
\begin{proof}
	Let $x\in C^{\perp_H},$ then $\langle x,c\rangle_H=0$  for all $c\in C.$ Then $\langle x ,c\rangle_{TrH}=Tr(\langle x,c\rangle_H)=Tr(0)=0$ for all $c\in C$ implies that $x\in C^{\perp_{TrH}}. $ Hence $C^{\perp_H}\subseteq C^{\perp_{TrH}}.$ Since $|C^{\perp_H}|=|C^{\perp_{TrH}}|$ therefore $C^{\perp_H}=C^{\perp_{TrH}}.$ Similarly, $C^{\perp}=C^{\perp_{TrE}}.$
\end{proof}

\section{Trace Hermitian ACD Codes}\label{trhacd}
In this section, we investigate trace Hermitian ACD codes over $\mathbb{F}_{q^2}.$ The following theorem provides a well-known characterization of linear complementary dual (LCD) codes over finite fields using generator matrices.

\begin{theorem}\cite[Proposition 1]{masseylcd}\cite[Proposition 1]{lcdhg}\label{lcdgenerator}
	Let $C$ be a linear code over a finite field $\mathbb{F}_q$ with a generator matrix $G$ of order $k\times n.$ Then $C$ is an Euclidean LCD code and a Hermitian LCD code if and only if $GG^T$ and $G\overline{G}^T$ are invertible matrices, respectively.
\end{theorem}

Every linear code can be viewed as an additive code. This insight leads us to the following results.
\begin{proposition}\label{hlcdtotracd}
	If $C$ is an $[n,k,d]_{q^2}$ Hermitian LCD code over  $\mathbb{F}_{q^2}$ then $C$ is a trace Hermitian ACD code over  $\mathbb{F}_{q^2}$ with parameters $(n,q^{2k},d)_{q^2}$.
\end{proposition}
\begin{proposition}
	Assume that $C$ is an $[n,k,d]_{q}$ LCD code over $\mathbb{F}_q$ with a generator matrix $G$. Suppose $D$ is a linear code over  $\mathbb{F}_{q^2}$  generated by the matrix $G.$ Then $D$ is an $(n,q^{2k},d)_{q^2}$  trace Hermitian ACD code when considered as an additive code over $\mathbb{F}_{q^2}.$
\end{proposition}
\begin{proof}
	Since $C$ is an LCD code over $\mathbb{F}_q$, therefore $GG^T$ is invertible over $\mathbb{F}_q.$ Also, $\overline{G}=G$ (since $G$ is matrix over $\mathbb{F}_q$) implies that $G\overline{G}^T$ is invertible over $\mathbb{F}_{q^2}.$ Therefore, by Theorem \ref{lcdgenerator}, $D$ is an LCD $[n,k,d]$ code with respect to Hermitian inner product  over $\mathbb{F}_{q^2}.$ Hence, by Proposition \ref{hlcdtotracd}, D is a trace Hermitian ACD $(n,q^{2k},d)_{q^2}$ code over $\mathbb{F}_{q^2}.$  
\end{proof}
Now, we characterize the elements in the trace Hermitian dual of  an additive code in connection with the generator matrix.
\begin{proposition}\label{xinperp}
	Assume that $G$ is a generator matrix for an additive $(n,q^k)_{q^2}$ code $C$ over $\mathbb{F}_{q^2}$. Then $x\in C^{\perp_{TrH}} $ if and only if $x\overline{G}^T+\overline{x}G^T=0.$
\end{proposition}
\begin{proof}
	Let $\{r_1,r_2,\dots,r_t\}$ be rows of $G.$ Then for $x\in \mathbb{F}_{q^2}^n,$
	$  x\overline{G}^T+\overline{x}G^T=0\\
	\iff  (x\cdot \overline{r_1}^T,x\cdot \overline{r_2}^T,\dots,x\cdot \overline{r_t}^T)+ (\overline{x}\cdot r_1^T,\overline{x}\cdot r_2^T,\dots,\overline{x}\cdot r_t^T)=0\\
	\iff  (x\cdot \overline{r_1}^T+\overline{x}\cdot r_1^T,x\cdot \overline{r_2}^T+\overline{x}\cdot r_2^T,\dots,x\cdot \overline{r_t}^T+\overline{x}\cdot r_t^T)=0\\
	\iff  x\cdot \overline{r_i}^T+\overline{x}\cdot r_i^T=0 \ \text{ for all } 1\leq i\leq t,
	\iff  \langle x, r_i\rangle_{TrH}=0 \text{ for all } 1\leq i\leq t\\
	\iff x\in C^{\perp_{TrH}}.$
\end{proof}

\begin{definition}\cite{complementsubfield}\label{orthomap}
	Suppose $\mathcal{V}$ is a vector space 
	over the field $\mathbb{F}_{q}$ with a given inner product $\langle\cdot ,\cdot\rangle$. A linear map $T: \mathcal{V} \to \mathcal{V}$ is said to be an orthogonal projection corresponding to the given inner product $\langle \cdot, \cdot \rangle$ if
	\begin{enumerate}
		\item $T^2=T$,
		\item $\langle x,w\rangle =0$ for all $w\in Ker(T)$ and $x\in Img(T),$ where $Img(T)$ is the  image set of $T$ and $Ker(T)$ is  the kernel of  $T.$
	\end{enumerate}
\end{definition}
A linear map $T$ is said to be a trace Euclidean orthogonal map (resp. a trace Hermitian orthogonal) if $T$ is an orthogonal map with respect to the trace Euclidean inner product (resp. trace Hermitian inner product).
\begin{lemma}\cite{af4}\label{nondeg}
	Notations are as in Definition \ref{orthomap}. If a given inner product is a non-degenerate inner product, then $Ker(T)$ is equal to the dual of $Img(T)$ with respect to the given inner product.      
\end{lemma}
Next, we examine the complementary dual code of additive codes and present two pivotal lemmas. Although these lemmas' proofs are analogous to those provided in \cite{complementsubfield}, we include them for completeness.
\begin{lemma} \label{tisprojection}
	Let $C$ be an additive code of length $n$ over $\mathbb{F}_{q^2},$ and $T:\mathbb{F}_{q^2}^n\to C $  be an $\mathbb{F}_p$-linear map. Then $T$ is a trace Hermitian  orthogonal projection onto $C$ if and only if \begin{equation*}
		T(v)=\begin{cases}
			v\ \ \text{ if } v\in C \\
			0 \ \ \text{ if }v\in C^{\perp_{TrH}}.\\
		\end{cases}
	\end{equation*} 
\end{lemma}
\begin{proof}
	Let $T$ be a trace Hermitian orthogonal projection onto $C$. Let $x\in C$ and $w\in C^{\perp_{TrH}}.$ Since $T$ is onto $C,$ there exists $u\in \mathbb{F}_{q^2}^n$ such that $T(u)=x$ and $x=T^2(u)=T(T(u))=T(x).$ By Lemma \ref{nondeg},  $w\in Ker(T)$, that is, $T(w)=0.$
	
	Conversely, suppose 
	\begin{equation*}
		T(v)=\begin{cases}
			v\ \ \text{ if } v\in C \\
			0 \ \ \text{ if }v\in C^{\perp_{TrH}}.\\
		\end{cases}
	\end{equation*}
	By the definition of $T$, we have $C\cap C^{\perp_{TrH}}=\{0\},$ which implies that $\mathbb{F}_{q^2}^n=C\oplus C^{\perp_{TrH}}.$ So, we can write each element $v$ in $\mathbb{F}_{q^2}^n$ as $v=u+w,$ where $u\in C$ and $w\in C^{\perp_{TrH}}.$  For $v\in\mathbb{F}_{q^2}^n ,$ $T(v)=T(u+w)=T(u)+T(w)=T(u)$ and $T^2(v)=T^2(u)=T(u)=T(v)$, that is, $T^2=T.$ Since $Img(T)=C$ and $Ker(T)=C^{\perp_{TrH}},$ $T$ is a trace Hermitian orthogonal projection onto $C.$
\end{proof}

\begin{lemma}\label{orthogonal}
	An additive code $C$ of length $n$ over $\mathbb{F}_{q^2}$ is a trace Hermitian ACD code if and only if there exists an $\mathbb{F}_p$-linear trace Hermitian orthogonal projection map from $\mathbb{F}_{q^2}^n$ onto $C.$
\end{lemma}
\begin{proof}
	Suppose $\Pi$ is an $\mathbb{F}_p$-linear trace Hermitian orthogonal projection  from $\mathbb{F}_{q^2}^n$ onto $ C.$ By Lemma \ref{tisprojection}, we have 
	\begin{equation*}
		\Pi(v)=\begin{cases}
			v\ \ \text{ if } v\in C \\
			0 \ \ \text{ if }v\in C^{\perp_{TrH}}\\
		\end{cases}
	\end{equation*} 
	for all $v\in \mathbb{F}_{q^2}^n.$ Let $x\in C\cap C^{\perp_{TrH}}.$ Then $\Pi(x)=x$ as $x\in C,$ and $\Pi(x)=0$ as $x\in C^{\perp_{TrH}}$ implies that $x=0.$ Therefore, $ C\cap C^{\perp_{TrH}}=\{0\}$, that is, $C$ is a trace Hermitian ACD code.
	
	Conversely, suppose $C$ is a trace Hermitian ACD code. Then $\mathbb{F}_{q^2}^n=C\oplus C^{\perp{TrH}},$ that is, for $v\in \mathbb{F}_{q^2}^n,$ there exist $x\in C$ and $y \in C^{\perp_{TrH}}$ such that $v=x+y.$ Define a map $\Pi: \mathbb{F}_{q^2}^n \to C $ as $\Pi(v)=x.$ Then $\Pi$ is an $\mathbb{F}_p$-linear map and 
	\begin{equation*}
		\Pi(v)=\begin{cases}
			v\ \ \text{ if } v\in C \\
			0 \ \ \text{ if }v\in C^{\perp_{TrH}}.\\
		\end{cases}
	\end{equation*} 
	By Lemma \ref{tisprojection}, $\Pi$ is an $\mathbb{F}_p$-linear trace Hermitian  orthogonal projection onto $C.$ 
\end{proof}

The following theorem establishes a criterion for identifying trace Hermitian additive complementary dual (ACD) codes over $\mathbb{F}_{q^2}$ using generator matrices. This criterion provides a systematic approach for the identification and construction of such codes by analyzing the structure of their generator matrices.
\begin{theorem} \label{hermitian acd}
	Suppose $G$ is a generator matrix of an $(n,q^k)_{q^2}$ additive code $C$ over $\mathbb{F}_{q^2}.$ Then $C$ is  a trace Hermitian ACD  code if and only if $G\overline{G}^T+\overline{G}G^T$ is an invertible matrix.
\end{theorem}
\begin{proof}
	If $G\overline{G}^T+\overline{G}G^T$ is not an invertible matrix, then there exists $0\neq u\in \mathbb{F}^t_{p}$ such that
	$u(G\overline{G}^T+\overline{G}G^T)=0,$ that is,    $(uG)\overline{G}^T+(\overline{uG})G^T=0$ (since $\overline{u}=u$). By Proposition  \ref{xinperp}, $uG\in C^{\perp_{TrH}}.$   Also, $uG\in C$; therefore, $uG \in C\cap C^{\perp_{TrH}}\neq \{0\}.$ Hence, $C$ is not a trace Hermitian ACD code. Conversely, suppose $G\overline{G}^T+\overline{G}G^T$ is invertible. Define a map  $\Pi: \mathbb{F}_{q^2}^n \to C$ as
	$$\Pi(v)=(v\overline{G}^T+\overline{v}G^T)(G\overline{G}^T+\overline{G}G^T)^{-1}G. $$
	We claim that the map $\Pi$ is an $\mathbb{F}_p$-linear trace Hermitian orthogonal projection from $\mathbb{F}_{q^2}^n$ onto $C.$ 
	Let $v\in C$, then there exists $u\in\mathbb{F}_p^t $ such that $v=uG.$
	\begin{align*}
		\Pi(v)=&(v\overline{G}^T+\overline{v}G^T)(G\overline{G}^T+\overline{G}G^T)^{-1}G\\
		=&(uG\overline{G}^T+\overline{uG}G^T)(G\overline{G}^T+\overline{G}G^T)^{-1}G\\
		=&u(G\overline{G}^T+\overline{G}G^T)(G\overline{G}^T+\overline{G}G^T)^{-1}G\  \ \text{ ( since $ \overline{u}=u$)}\\
		=&uG=v.
	\end{align*} 
	If $v\in C^{\perp_{TrH}},$ then by Proposition \ref{xinperp}, $v\overline{G}^T+\overline{v}G^T=0.$ This implies that $\Pi(v)=(v\overline{G}^T+\overline{v}G^T)(G\overline{G}^T+\overline{G}G^T)^{-1}G=0$. According to Lemma \ref{tisprojection}, $\Pi$ is an $\mathbb{F}_p$-linear trace Hermitian orthogonal projection from $\mathbb{F}_{q^2}^n$ onto $C$, consequently, by Lemma \ref{orthogonal}, $C$ is a trace Hermitian ACD code.
\end{proof}
We say any two elements $a, b$ in $\mathbb{F}_{q^2}$ are \textit{trace Euclidean  orthogonal} and  \textit{trace Hermitian  orthogonal} if $\langle a,b\rangle_{TrE}=Tr(ab)=0$ and $\langle a,b\rangle_{TrH}=Tr(a\overline{b})=0$, respectively. The elements $a$ and $b$ are said to be linearly independent over $\mathbb{F}_q$ if for $\alpha, \alpha' \in \mathbb{F}_q$, $\alpha a+\alpha' b=0$ implies $\alpha=\alpha'=0.$
\begin{lemma}
	There exist nonzero $a,b\in \mathbb{F}_{q^2}$ such that $a$ and $b$ are trace Hermitian orthogonal, that is, $Tr(a\overline{b})=0.$
\end{lemma}
\begin{proof}
	Let $\zeta$ be a primitive element of $\mathbb{F}_{q^2}$. Take $a=\zeta$ and $b=\zeta^{-q+1}-1,$ then $\overline{b}=(\zeta^{-q+1}-1)^q=\zeta^{q-1}-1,$ and $Tr(a\overline{b})=a\overline{b}+\overline{a}b=\zeta \cdot (\zeta^{q-1}-1)+ \zeta^q\cdot (\zeta^{-q+1}-1)=0.$ 
\end{proof}
\begin{theorem}\label{abcdh}
	Let $C$ and $D$ be LCD codes over $\mathbb{F}_q$ with parameters $[n,k_1]_{q}$ and $[n,k_2]_{q},$ respectively. Let $ a,b\in \mathbb{F}_{q^2}$ be nonzero trace Hermitian orthogonal elements in $\mathbb{F}_{q^2}$. Then the code $E=aC+bD$ is an additive code over  $\mathbb{F}_{q^2}$ with $E^{\perp_{TrH}}=aC^{\perp_E}+bD^{\perp_E}$. Moreover, if $a$ and $b$ are linearly
	independent over $\mathbb{F}_q,$ then $E$ is a trace Hermitian ACD $(n,q^{k_1+k_2})_{q^2}$ code over $\mathbb{F}_{q^2}.$
\end{theorem}
\begin{proof}
	Let $x\in aC^{\perp_E}+bD^{\perp_E} .$ Then there are $c'\in C^{\perp}$ and $d'\in D^{\perp}$ such that $x=ac'+bd'.$ For $y=ac+bd\in E,$ where $c\in C$ and $d\in D$, we have
	\begin{align*}
		\langle x,y\rangle_{TrH}=&\ \langle ac'+bd',ac+bd\rangle_{TrH}\\
		=&\ (ac'+bd')\cdot \overline{(ac+bd)}+(\overline{ac'+bd'})\cdot (ac+bd)\\
		=&\  a\overline{a}(c'\cdot c)+a\overline{b}(c'\cdot d)+b\overline{a}(d'\cdot c)+b\overline{b}(d'\cdot d) \\
		&\ +\overline{a}a(c\cdot c')+\overline{a}b(c'\cdot d)+\overline{b}a(d'\cdot c)+\overline{b}b(d\cdot d')\\
		=&\ (a\overline{b}+\overline{a}b)(c'\cdot d)+(a\overline{b}+\overline{a}b)(d'\cdot c)\\
		=&\ Tr(a\overline{b})(c'\cdot d+d'\cdot c) =0.
	\end{align*}
	This implies that $aC^{\perp_E}+bD^{\perp_E} \subseteq E^{\perp_{TrH}}.$ Also, $|E^{\perp_{TrH}}|=\frac{q^{2n}}{|E|}=q^{2n-k_1-k_2}$ which is equal to $|aC^{\perp_E}+bD^{\perp_E}|=q^{2n-k_1-k_2}.$ Therefore, $E^{\perp_{TrH}}=aC^{\perp_E}+bD^{\perp_E}.$ Let $x\in E\cap E^{\perp_{TrH}}$. Then there exist $c\in C, d\in D$ and $c'\in C^{\perp_{E}},\ d'\in D^{\perp_{E}}$ such that $x=ac+bd=ac'+bd'. $ If $a$ and $b$ are linearly independent over $\mathbb{F}_q,$ then $c=c'$ and $d=d'.$ This implies that $c\in C\cap C^{\perp_E}$ and $d\in D\cap D^{\perp_E}.$ Since $C$  and $D$ are LCD codes, $c=d=0$, and consequently, $x=0.$ Hence, $E$ is a trace Hermitian ACD  $(n,q^{k_1+k_2})_{q^2} $ code over $\mathbb{F}_{q^2}.$
\end{proof}

\begin{proposition}
	Let $G_C$ and $G_D$ be generator matrices for linear codes $C$ and $D$ over $\mathbb{F}_q,$  respectively. Then $G_E=\begin{bmatrix}
		aG_C\\
		bG_D
	\end{bmatrix}$ is a generator matrix for the additive code $E=aC+bD$  over $\mathbb{F}_{q^2}$ where  $a,b\in \mathbb{F}_{q^2}$  such that $a$ and $b$ are linearly independent over  $\mathbb{F}_q.$
\end{proposition}
\begin{proof}
	By  Theorem \ref{abcdh}, $dim_{\mathbb{F}_q}(E)=k_1+k_2.$ So, it is sufficient to show that the rows of $G_E$ are linearly independent over $\mathbb{F}_q.$  Let $\{r_1,r_2,\dots,r_{k_1}\}$ and $\{m_1,m_2,\dots, m_{k_2}\}$ be rows of matrices $G_C$ and $G_D,$ respectively. Then rows of matrix $G_E$ are $\{ar_1,ar_2,\dots,ar_{k_1}, bm_1,bm_2,\dots, bm_{k_2}\}$.  Let $s_1ar_1+s_2ar_2+\dots+s_{k_1}ar_{k_1}+t_1bm_1+t_2bm_2+\dots+t_{k_2}bm_{k_2}=0$ for $s_i,t_j\in \mathbb{F}_q,\ 1\leq i\leq k_1,\ 1\leq j\leq k_2.$ This implies that $a[s_1r_1+\dots +s_{k_1}r_{k_1}]+b[t_1m_1+\dots+t_{k2}m_{k_2}]=0.$ Since $a$ and $b$ are linearly independent over  $\mathbb{F}_q,$ therefore, $s_1r_1+\dots +s_{k_1}r_{k_1}=0$ and $t_1m_1+\dots+t_{k_2}m_{k_2}=0.$ Also, $\{r_1,r_2,\dots,r_{k_1}\}$ and $\{m_1,m_2,\dots, m_{k_2}\}$ are linearly independent over $\mathbb{F}_q$. Hence, $s_i=0=t_j$ for $1\leq i\leq k_1,\ 1\leq j\leq k_2.$ This completes the proof.
\end{proof}
\begin{example}
	Let $C$  be a linear $[7,2]_3$ code over $\mathbb{F}_{3}$ with a generator matrix 
	$$G_C=\begin{bmatrix}
		1& 0& 2&1& 2& 2& 1\\
		0& 1& 0& 2& 1& 2& 2
	\end{bmatrix},$$
	and $D$ be another $[7,3]_3$ linear code over $\mathbb{F}_{3}$ with a generator matrix $$ G_D=\begin{bmatrix}
		1& 0& 0& 2& 2& 1& 2\\
		0 &1& 1& 2& 1& 2& 2\\
		1 &0& 1 &2& 0 &2 &2\\
	\end{bmatrix}.$$
	Clearly, $C$ and $D$ are LCD codes over  $\mathbb{F}_{3}.$ Take $a=\beta$ and $b=2\beta+1$ in $\mathbb{F}_{9}=\{0,1,2,\beta,\beta +1,\beta +2,2\beta,2\beta +1,2\beta+2\},$ where $\beta^2+2\beta+2=0.$ Then $ Tr(a\overline{b})=0$ and $a,b$ are linearly independent over $\mathbb{F}_3.$   By Proposition \ref{abcdh}, the additive code $E=aC+bD$ is a trace Hermitian $(7,3^5)_{9}$ ACD code over $\mathbb{F}_{9}$ with a generator matrix  
	$$ G_E=\begin{bmatrix}
		\beta& 0& 2\beta&\beta& 2\beta& 2\beta& \beta\\
		0 &2\beta+1& 2\beta+1& \beta+2& 2\beta+1& \beta+2& \beta+2\\
		2\beta+1& 0& 0& \beta+2& \beta+2& \beta+2\beta+1& \beta+2\\
		0& \beta& 0& 2\beta& \beta& 2\beta& 2\beta\\
		2\beta+1 &0& 2\beta+1 &\beta+2& 0 &\beta+2 &\beta+2\\
	\end{bmatrix}.$$
	
\end{example}

\section{Trace Euclidean ACD Codes}\label{treacd}
This section investigates additive complementary dual (ACD) codes over the field $\mathbb{F}_{q^2}$ with respect to the trace Euclidean inner product. We provide a comprehensive discussion of trace Euclidean ACD codes, and based on this description, we construct trace Euclidean ACD codes over $\mathbb{F}_{q^2}$ by employing linear complementary dual (LCD) codes over $\mathbb{F}_{q}.$ The following proposition provides an explanation of the elements in the trace Euclidean dual of an additive code using the generator matrices.
\begin{proposition}\label{xinperpe}
	Let $G$ be a generator matrix of an additive $(n,q^k)_{q^2}$ code $C$ over $\mathbb{F}_{q^2}$. Then $x\in C^{\perp_{TrE}} $ if and only if $xG^T+\overline{x}\ \overline{G}^T=0.$
\end{proposition}
\begin{proof}
	Let $\{\rho_1,\rho_2,\dots,\rho_t\}$ be rows of $G.$ Then
	$xG^T+\overline{x}\overline{G}^T=0\\
	\iff (x\cdot\rho_1^T ,x\cdot \rho_2^T ,\dots,x\cdot \rho_t^T )+ (\overline{x}\cdot \overline{\rho_1}^T,\overline{x}\cdot \overline{\rho_2}^T,\dots,\overline{x}\cdot \overline{\rho_t}^T)=0\\
	\iff  (x\cdot \rho_1^T +\overline{x}\cdot \overline{\rho_1}^T,x\cdot \rho_2^T +\overline{x}\cdot \overline{\rho_2}^T,\dots,x\cdot \rho_t^T+\overline{x}\cdot \overline{\rho_t}^T )=0\\
	\iff  x\cdot \rho_i^T +\overline{x}\cdot \overline{\rho_i}^T =0 \ \text{ for all } 1\leq i\leq t\\
	\iff  \langle x, \rho_i\rangle_{TrE}=0 \text{ for all } 1\leq i\leq t\\
	\iff x\in C^{\perp_{TrE}}.$
\end{proof}
%The proof of following lemmas are similar to Lemma \ref{tisprojection} and Lemma \ref{orthogonal}.
\begin{lemma} 
	Let $C$ be an additive code of length $n$ over $\mathbb{F}_{q^2}$ and $T:\mathbb{F}_{q^2}^n\to C $  be an $\mathbb{F}_p$-linear map. Then $T$ is a trace Euclidean orthogonal projection onto $C$ if and only if \begin{equation*}
		T(v)=\begin{cases}
			v\ \ \text{ if } v\in C \\
			0\ \  \text{ if }v\in C^{\perp_{TrE}}.\\
		\end{cases}
	\end{equation*} 
\end{lemma}

\begin{lemma}\label{iffprojectio euclidean}
	An additive code $C$ of length $n$ over $\mathbb{F}_{q^2}$ is a trace Euclidean ACD code  if and only if there exists an $\mathbb{F}_p$-linear trace Euclidean orthogonal projection map from $\mathbb{F}_{q^2}^n$ onto $C.$
\end{lemma}

%The trace Euclidean ACD codes in terms of generator matrices has been characterized in the following theorem.
Next theorem characterizes trace Euclidean ACD codes in terms of generator matrix.
\begin{theorem}\label{euclidean acd}
	Let $G$ be a generator matrix for an $(n,q^k)_{q^2}$ additive code $C$ over $\mathbb{F}_{q^2}.$ Then $C$ is a trace Euclidean ACD code if and only if $G\ G^T+\overline{G} \ \overline{G}^T$ is an  invertible matrix.
\end{theorem}
\begin{proof}
	The proof follows similar to the proof of Theorem \ref{hermitian acd}.  The trace Euclidean orthogonal projection onto $C$ is given by the map $\Pi: \mathbb{F}_{q^2}^n \to C$, defined as
	$\Pi(v)=(v G^T+\overline{v}\overline{G}^T)(GG^T+\overline{G}\ \overline{G}^T)^{-1}G.$
\end{proof}

\begin{lemma}
	There exist nonzero $ a,b\in \mathbb{F}_{q^2}$ such that $a$ and $b$ are trace Euclidean orthogonal, that is, $Tr(ab)=0.$
\end{lemma}
\begin{proof}
	Let $\zeta$ be a generator of multiplicative group $\mathbb{F}_{q^2}\setminus \{0\}$. Take $a=\zeta$ and $b=\zeta^{q-1}-1$ in $\mathbb{F}_{q^2}.$  Then $Tr(ab)=Tr(\zeta^q-\zeta)=\zeta^q-\zeta +(\zeta^q-\zeta)^q=0.$ 
\end{proof}

\begin{theorem}
	Let $C$ and $D$ be LCD $[n,k_1]_q$ and $[n,k_2]_q$ codes over $\mathbb{F}_q$. Let $a,b\in \mathbb{F}_{q^2}$ be nonzero trace Euclidean orthogonal elements, that is, $Tr(ab)=0.$ If $a$ and $b$ are linearly independent over $\mathbb{F}_q$,  then the code $E=aC+bD$ is a trace Euclidean ACD $(n,q^{k_1+k_2})_{q^2}$ code over $\mathbb{F}_{q^2}$.
\end{theorem}
\begin{proof}
	First, we show that $E^{\perp_{TrE}}=aC^{\perp_E}+bD^{\perp_E}.$ Let $x\in aC^{\perp_E}+bD^{\perp_E}$. Then there exist $c'\in C^{\perp_E}$ and $d'\in D^{\perp_E}$ such that $x=ac'+bd'.$ For any $y= ac+bd\in E$, where $c\in C$ and $d\in D$, we have
	\begin{align*}
		\langle x,y\rangle_{TrE}=&\langle ac'+bd',ac+bd\rangle_{TrE}\\
		=& (ac'+bd')(ac+bd)+\overline{(ac'+bd')(ac+bd)}\\
		=& aa (c'\cdot c)+ab (c'\cdot d)+ba(d'\cdot c)+bb(d'\cdot d) \\
		&\ + \overline{aa} (c'\cdot c)+\overline{ab} (c'\cdot d)+\overline{ba}(d'\cdot c)+\overline{bb}(d'\cdot d)\\
		=& (ab+\overline{ab}) (c'\cdot d)+(ba+\overline{ba})(d'\cdot c)\\
		=& Tr(ab)(c'\cdot d+d'\cdot c)=0. 
	\end{align*}
	This implies that $aC^{\perp_E}+bD^{\perp_E}\subseteq E^{\perp_{TrE}}.$ Observe that $|E|=q^{k_1+k_2},$ and  $|aC^{\perp_E}+bD^{\perp_E}|=q^{2n-(k_1+k_2)}.$ Hence $E^{\perp_{TrE}}=aC^{\perp_E}+bD^{\perp_E}.$ 
	Let $x\in E\cap E^{\perp_{TrE}}$, then there exist $c\in C,d\in D$ and $c'\in C^{\perp_E},d'\in D^{\perp_E}$ such that $x=ac+bd=ac'+bd'.$ Since $a$ and $b$ are linearly independent over $\mathbb{F}_q,$ we get $c=c'$ and $d=d'$. Consequently,  $c\in C\cap C^{\perp_E}$ and $d\in D\cap D^{\perp_E}$. Also, $C$ and $D$ are LCD codes over $\mathbb{F}_q,$  implies that $c=0$ and $d=0,$ that is, $x=0.$ Hence, $E$ is trace Euclidean $(n,q^{k_1+k_2})_{q^2}$ ACD code.
\end{proof}

\begin{example}
	Suppose $G_C$ and $G_D$ are generator matrices for LCD codes $C$ and $D$ over $\mathbb{F}_5$ with parameters $[6,3,3]_5$ and $[6,2,4]_5,$ respectively, where 
	$$G_C=\begin{bmatrix}
		0&1&0&1&2&0\\
		0&2&3&0&2&4\\
		1&0&3&4&2&1
	\end{bmatrix},\ G_D=\begin{bmatrix}
		0&2&3&0&2&4\\
		1&1&0&2&4&0
	\end{bmatrix}.$$
	Let $a=\omega $ and $b=2\omega +1$ be in $\mathbb{F}_{25}=\mathbb{F}_5[\omega],$ where $\omega^2+4\omega+2=0.$ Then additive code $E=aC+bD$ is a trace Euclidean ACD $(6,5^5,2)_{25}$ code  with a generator matrix
	$$G_E=\begin{bmatrix}
		0 &\omega &0 &\omega & 2\omega& 0\\
		0& 2 \omega &3\omega &0&2\omega &4\omega\\
		\omega &0&3\omega&4\omega& 2\omega&\omega\\
		0&4\omega+2&\omega+3&0&4\omega+2&3\omega+4\\
		2\omega+1&2\omega+1&0&4\omega+2&3\omega+4&0
	\end{bmatrix}.$$
\end{example}

\section{Construction of trace Euclidean and trace Hermitian ACD codes}\label{construction acd}
The description of ACD codes in terms of generator matrices provided in Theorems \ref{hermitian acd} and \ref{euclidean acd} plays an important part in the construction of ACD codes. In \cite{verma2023}, authors constructed Galois LCD codes with various parameters from a given Galois LCD code over finite fields. Similarly,  we give methods for the construction of ACD codes with  parameters $(n+1,q^{k+1})_{q^2}$ and $(n,q^{k+1})_{q^2}$  from a given  $(n,q^k)_{q^2}$ ACD code corresponding to the trace inner products. 

\begin{theorem}\label{conshermitian}
	Assume that $\mathcal{C}$ is an $(n,q^k)_{q^2}$ ACD code  with respect to the trace Hermitian inner product and $G$ is a  generator matrix for $\mathcal{C}.$  Let $u\in \mathcal{C}^{\perp_{TrH}}.$ 
	\begin{enumerate}[(i)]
		\item If $q$ is odd and $\langle u,u\rangle_{TrH}=0,$ then the additive code $\mathcal{C}_1$ generated by the matrix 
		$$G_1=\begin{bmatrix}
			1 & u\\
			0 & G
		\end{bmatrix}$$
		is an $(n+1,q^{k+1})_{q^2}$  trace Hermitian ACD code.
		\item  If $\langle u,u\rangle_{TrH}\neq 0,$ then the additive code $\mathcal{C}_2$ generated by the matrix 
		$$G_2=\begin{bmatrix}
			u\\
			G
		\end{bmatrix}$$
		is an $(n,q^{k+1})_{q^2}$  trace Hermitian ACD code.
		
	\end{enumerate}
\end{theorem}

\begin{proof}
	Since $u\in \mathcal{C}^{\perp_{TrH}}$, by Proposition \ref{xinperp}, we have $u\overline{G}^T+\overline{u}G^T=0.$ This implies that $G\overline{u}^T+\overline{G}u^T=0.$ Also, $C$ is a trace Hermitian ACD code. Therefore, by Theorem \ref{hermitian acd}, $G\overline{G}^T+ \overline{G}G^T$ is an invertible matrix. 
	\begin{enumerate}[(i)]
		\item Let $\langle u,u\rangle_{TrH}=0$ and $q$ be odd. Then
		\begin{align*}
			G_1 \overline{G_1}^T+\overline{G_1}G_1^T=&\begin{bmatrix}
				1 & u\\
				0 & G
			\end{bmatrix}
			\begin{bmatrix}
				1 & 0 \\
				\overline{u}^T & \overline{G}^T
			\end{bmatrix}+\begin{bmatrix}
				1 & \overline{u}\\
				0 & \overline{G}
			\end{bmatrix}
			\begin{bmatrix}
				1 & 0\\
				u^T & G^T
			\end{bmatrix}\\
			=&\begin{bmatrix}
				1+u\overline{u}^T & u\overline{G}^T\\
				G\overline{u}^T & G\overline{G}^T
			\end{bmatrix}+
			\begin{bmatrix}
				1+\overline{u}u^T &\overline{u}G^T \\
				\overline{G}u^T & \overline{G}G^T
			\end{bmatrix}\\
			=&\begin{bmatrix}
				2+\langle u,u\rangle_{TrH} & u\overline{G}^T+\overline{u}G^T \\
				G\overline{u}^T+\overline{G}u^T &G\overline{G}^T+ \overline{G}G^T
			\end{bmatrix}\\
			=&\begin{bmatrix}
				2 & 0\\
				0 &G\overline{G}^T+ \overline{G}G^T
			\end{bmatrix}.
		\end{align*} 
		
		Thus, $G_1 \overline{G_1}^T+\overline{G_1}G_1^T$
		is an invertible matrix. Hence $C_1$ is a trace Hermitian ACD code.	
		
		\item Let $\langle u,u\rangle_{TrH}\neq 0.$ Then 
		\begin{align*}
			G_2 \overline{G_2}^T+\overline{G_2}G_2^T=&\begin{bmatrix}
				u\\
				G
			\end{bmatrix}
			\begin{bmatrix}
				\overline{u}^T & \overline{G}^T
			\end{bmatrix}+\begin{bmatrix}
				\overline{u}\\
				\overline{G}
			\end{bmatrix}
			\begin{bmatrix}
				u^T & G^T
			\end{bmatrix}\\
			=&\begin{bmatrix}
				u\overline{u}^T & u\overline{G}^T\\
				G\overline{u}^T & G\overline{G}^T
			\end{bmatrix}+
			\begin{bmatrix}
				\overline{u}u^T &\overline{u}G^T \\
				\overline{G}u^T & \overline{G}G^T
			\end{bmatrix}\\
			=&\begin{bmatrix}
				\langle u,u\rangle_{TrH} & u\overline{G}^T+\overline{u}G^T \\
				G\overline{u}^T+\overline{G}u^T &G\overline{G}^T+ \overline{G}G^T
			\end{bmatrix}\\
			=&\begin{bmatrix}
				\langle u,u\rangle_{TrH} & 0\\
				0 &G\overline{G}^T+ \overline{G}G^T
			\end{bmatrix}.	
		\end{align*}
		Thus, $G_2 \overline{G_2}^T+\overline{G_2}G_2^T$ is an invertible matrix. Hence $C_2$ is a trace Hermitian ACD code.
	\end{enumerate}	
\end{proof}

In the following theorem, we give a construction method for trace Euclidean ACD codes. We omit the proof since it follows on similar lines of proof from Theorem \ref{conshermitian}. 
\begin{theorem}\label{conseuclidean}
	Assume that $\mathcal{C}$ is an $(n,q^k)_{q^2}$ ACD code  with respect to the trace Euclidean inner product and $G$ is a  generator matrix for $\mathcal{C}.$  Let $v\in \mathcal{C}^{\perp_{TrE}} .$ 
	\begin{enumerate}[(i)]
		\item If $q$ is odd and $\langle v,v\rangle_{TrE}=0,$ then the additive code $\mathcal{C}_1$ generated by the matrix 
		$$G_1=\begin{bmatrix}
			1 & v\\
			0 & G
		\end{bmatrix}$$
		is an $(n+1,q^{k+1})_{q^2}$  trace Euclidean ACD code.
		\item  If $\langle v,v\rangle_{TrE}\neq 0,$ then the additive code $\mathcal{C}_2$ generated by the matrix 
		$$G_2=\begin{bmatrix}
			v\\
			G
		\end{bmatrix}$$
		is an $(n,q^{k+1})_{q^2}$  trace Euclidean ACD code.
	\end{enumerate}
\end{theorem}
Using Theorem \ref{conseuclidean} and  \ref{conshermitian}, we construct several ACD codes with better parameters than those available in the literature.  First, we search  $(n,q^k)_{q^2}$ trace Euclidean (resp. trace Hermitian) ACD codes with the help of MAGMA \cite{magma} functions {\fontfamily{qcr}\selectfont
	BKLC} or  {\fontfamily{qcr}\selectfont
	RandomAdditiveCode}. Then repeatedly apply  Theorem \ref{conseuclidean} (resp. Theorem \ref{conshermitian}), we get new trace Euclidean (resp. trace Hermitian) ACD codes. All computations are performed in MAGMA\cite{magma}, and we compare the codes with Grassl's code  table \cite{codetable} and the MAGMA database. 
\begin{example}
	Using  {\fontfamily{qcr}\selectfont
		RandomAdditiveCode}, we found a $(15,2^4,11)_4$ trace Euclidean ACD code over $\mathbb{F}_4=\mathbb{F}_2[w]$ with a generator matrix 	
	$$ \begin{bmatrix}
		1&  w&   w&   0&    w&    1&  w^2&    w&  w^2&   0&  w^2&    w&      w&    1&   1\\
		w&  1&   0&   w&   w^2&   1&    w&  w^2&    1&   0&    1&    w&      w&  w^2& w^2\\
		0&  0&  w^2&  w&   w^2&   1&  w^2&   0&   w^2&   1&    0&   w^2&     1&  w^2&   1\\
		0&  0&   0&   1&     w&   w&    w&   0&     1&   w&  w^2&     1&   w^2&    1&   w
	\end{bmatrix},$$
	where $w^2+w+1=0.$ Applying Theorem \ref{conseuclidean} for a vector $$v=(  1, 1, 1, 1, w, 1, 0, 0, w, 1, 0, 0, 0, w, 1),$$ we obtained  $(15,2^5,10)_4$ trace Euclidean ACD code over $\mathbb{F}_4.$ 
	The best known binary code of length $15$ and size $2^5$ has the minimum distance $7$. In contrast, the constructed code over $\mathbb{F}_4$, has the same length and size but a significantly large minimum distance $10.$   
\end{example}

\begin{example}
	By random search in MAGMA \cite{magma}, we obtained $(15,2^6,11)_4$ trace Euclidean ACD code over $\mathbb{F}_4=\mathbb{F}_2[w]$ with generator matrix 
	$$\begin{bmatrix}
		1 & 0 & 0 & w & w^2 & w & w & w & w & w^2 & w & 0 & 0 & 1 & w^2\\
		0& 1& 0& w& w^2& w^2& w^2& 0& w& w& 0& w^2& w^2& w^2& w\\
		0& 0& 1& 0& w^2& w& 1& w^2& 1& 0& 1& w^2& 1& w& w^2\\
		w& 0& 0& w^2& w^3& w^2& w^2& w^2& w^2& w^3& w^2& 0& 0& w& w^3\\
		0& w& 0& w^2& w^3& w^3& w^3& 0& w^2& w^2& 0& w^3& w^3& w^3& w^2\\
		0& 0& w& 0& w^3& w^2& w& w^3& w& 0& w& w^3& w& w^2& w^3
	\end{bmatrix},$$
	where $w^2+w+1=0.$ Applying Theorem \ref{conseuclidean} with a vector $$v=(w^2, w^2,   0, w^2,   0,  w,   1,   1,   0,   0,   0,   0,   1,   w,   1),$$ we get $(15,2^7,9)_4$ trace Euclidean ACD code.	
\end{example}

\begin{example}
	Using  {\fontfamily{qcr}\selectfont
		RandomAdditiveCode} function in MAGMA, we found a $(13,3^7,8)_9$ trace Euclidean ACD code over $\mathbb{F}_9=\mathbb{F}_3[w]$ with a generator matrix \\
	\begin{equation*}
		\resizebox{\hsize}{!}{$\begin{bmatrix}
				1 & 0&1&0&w&2&1&0&1&2&2&1&0\\
				w&1& 0& 2w& 2w& w& 1& 0& w + 1& 2w + 2& 1& 2&2\\
				w& w& w& w& w+1& w& w& w& w& w& w&w+1&1\\
				0& 2w + 1& 2w& w& w& 2w + 1& w + 1& 2w + 1& w&w + 1& w + 1&0&w\\
				0& 0& 2w + 1& 2w& w& 2w + 2& 1& 0& w + 2& w + 1& 2w + 2&0&w+1\\
				0& w+1& 0& 2w+ 1& 2w& 1+w& w + 2& 2& 2w + 2& w + 1& 0&1&0\\
				0& 0& 0& 0& 2w + 1& 2w& 1& w + 2& 2& 2w + 2& w + 1&0&w
			\end{bmatrix}$},
	\end{equation*}
	where $w^2+1=0.$
	Through the recursive application of  Theorem \ref{conseuclidean} with appropriate choices of vector $v,$ we obtained trace Euclidean ACD codes with parameters
	\begin{align*}
		(13,3^8,8)_9,\ (13,3^9,8)_9,\ (14,3^9,8)_9,\ (14,3^{10},8)_9,\ (15,3^{11},8)_9\ (15,3^{12},8)_9.	
	\end{align*}
	
\end{example}

\begin{example}
	Using  {\fontfamily{qcr}\selectfont
		RandomAdditiveCode} function in MAGMA, we found trace Euclidean ACD codes over $\mathbb{F}_9$ with the following parameters
	\begin{align*}
		(11,3^6,3)_9,\ (13,3^{10},5)_9,\ (14,3^4,12)_9,\ (15,3^{10},7)_9,\ (17,3^{10},7)_9,\ (18,3^{10},8)_9,\ (20,3^{10},9)_9.\ 
	\end{align*} Using Theorem \ref{conseuclidean}, we constructed trace Euclidean ACD codes over $\mathbb{F}_9$ with the following parameters
	\begin{align*}
		(11,3^7,6)_9,\ (13,3^{11},5)_9,\ (14,3^5,11)_9,\ (16,3^{11},7)_9,\ (17,3^{11},7)_9,\\
		(18,3^{11},8)_9,\ (19,3^{11},8)_9,\ (20,3^{11},9)_9,\ (21,3^{11},9)_9.\ 
	\end{align*}
	We list the constructed trace Euclidean ACD codes with  appropriate choices of vector $v$ in Table \ref{teuclidean}, and we compare them  with the best known linear codes of the same length and size available in Grassl's code table \cite{codetable} and MAGMA \cite{magma} databases.
	
\end{example}

\begin{table}[h!]
	\begin{center}
%	\caption{ Trace Euclidean ACD codes over the finite field $\mathbb{F}_9=\mathbb{F}_3[w]$ constructed using Theorem \ref{conseuclidean}, where $w^2+1=0.$ \label{teuclidean} }
		\begin{adjustbox}{width=\textwidth}
			\begin{tabular}{|c|c|c|c|}
				\hline
				$C$&$v$&Constructed code& Best known linear code\\
				\hline
				$(11,3^6,6)$&$(2w+1,w+2,2w+2,w+2,0,0,0,w,0,0,2w)$&$(11,3^7,6)$& $(11,3^7,3)$\\
				\hline
				$(12,3^{4},10)$&$( 0,0, w^5, w^5, w^5, w^6, 1, 1, w^5,w^3, 2, 1) \ \  (\text{ here }w^2+2w+2=0)$&$(13,3^{5},10)$& $(13,3^{5},8)$\\
				\hline 
				$(13,3^7,8)$&$(w+1,w+2,2w+1,2,2w+2,w+2,0,0,0,2w,2w,w+2,2w+2)$&$(13,3^8,8)$& $(13,3^8,7)$\\
				\hline
				$(13,3^8,8)$&$(w + 1, w + 1, 2, 2w + 1, 2w, 2, w + 2, 0, 0, 0, 2w, 2, 2w + 1)$&$(13,3^9,8)$& $(13,3^9,4)$\\
				\hline
				$(13,3^8,8)$&$(w+1, 2w, 1, 0, 0,  0,  0,  0, 0,  w, 2, w,  2)$&$(14,3^9,8)$& $(14,3^9,5)$\\
				\hline
				$(13,3^{10},5)$&$(2w+2,2w+2,1,w+2,w,0,0,w,w+1,w,2w+1,0,w+2)$&$(13,3^{11},5)$& $(13,3^{11},2)$\\
				\hline
				$(14,3^{4},12)$&$ (1, w^2,   2,   w,   2, w^5,   1, w^5,   1,   0,   0,   0, w^3,   1)\ \  (\text{ here }w^2+2w+2=0)$&$(14,3^{5},11)$& $(14,3^5,9)$\\
				\hline
				$(14,3^{9},8)$&$(2w + 2, 1, 2, 1, 0,0, 0, 0, 0, 1, 0, 2, 2w + 2, w)
				$&$(14,3^{10},8)$& $(14,3^{10},7)$\\
				\hline
				$(14,3^{10},8)$&$(w+1, w+2, 0, 0, 0, 0, 0, 0, 1, 2, 2w, w, 2w+2, w + 1)
				$&$(15,3^{11},8)$& $(15,3^{11},4)$\\
				\hline
				
				$(15,3^{11},8)$&$(w+1,2w +2, w+1, 2w, w, w +2, 0, 0, 0, w, 1, 2w+1, 2, 2w +1, w)
				$&$(15,3^{12},8)$& $(15,3^{12},7)$\\
				
				%				\hline
				%				$(15,3^{10},7)$&$(0,2w+1,2,2,2,0,0,2w,0,w+2,w+2,1,1,0,w+1)$&$(15,3^{11},7)$& $(15,3^{11},3)$\\
				\hline
				$(15,3^{10},7)$&$(1,2w+1,2,0,w,0,0,w,0,2w,w+2,2,1,0,w)$&$(16,3^{11},7)$& $(16,3^{11},4)$\\
				\hline
				$(17,3^{10},7)$&$(w+1,w+2,2w+1,2w,w+1,1,0,0,0,0,w+1,1,w,0,w,0,2w+2)$&$(17,3^{11},7)$& $(17,3^{11},4)$\\
				\hline
				$(18,3^{10},8)$&$(2,2w,w,1,1,0,0,0,0,0,0,2,2,0,2w,0,w+2,0)$&$(18,3^{11},8)$& $(18,3^{11},5)$\\
				
				\hline
				$(18,3^{10},8)$&$(w+2,2w+1,w+1,1,1,0,0,0,0,1,w+2,2,2w,0,2w,0,1,0)$&$(19,3^{11},8)$& $(19,3^{11},6)$\\
				\hline
				$(20,3^{10},9)$&$(1,w,2w+2,2w+2,w,0,0,0,0,0,0,0,0,2w,0,0,0,w+1,w,2)$&$(20,3^{11},9)$& $(20,3^{11},6)$\\
				\hline
				$(20,3^{10},9)$&$(w,2w+2,2w+2,w,0,0,0,0,0,0,0,0,2w,0,0,0,w+1,w,2,w+2)$&$(21,3^{11},9)$& $(21,3^{11},7)$\\
				\hline
				%				$(13,3^{5},10)$&$ (u^5,   u, u, 2, u^5, 1, u^5, 1, 0, 0,   0, 0, u^6, u)$&$(14,3^{5},10)$& $(14,3^5,9)$\\
				%				\hline		
			\end{tabular}
		\end{adjustbox}
	\end{center}
\caption{Trace Euclidean ACD codes over finite field $\mathbb{F}_9=\mathbb{F}_3[w]$ constructed using Theorem \ref{conseuclidean}, where $w^2+1=0.$  \label{teuclidean}}
\end{table}

\begin{example}
	Through random search in  MAGMA software \cite{magma}, we obtained trace Hermitian ACD codes over $\mathbb{F}_9$ with the following parameters
	\begin{align*}
		(11,3^6,6)_9,\ (12,3^4,10)_9,\  (13,3^{7},8)_9,\ (13,3^{10},5)_9,\ (14,3^{4},12)_9,\\ (14,3^{8},8)_9,  (15,3^{11},8)_9,\ (18,3^{10},8,\ )_9 (20,3^{10},9)_9.
	\end{align*}
	
	By recursively applying Theorem \ref{conshermitian}, we constructed trace Hermitian ACD codes over $\mathbb{F}_9$ with parameters
	\begin{align*}
		(11,3^{7},6)_9,\ 	(13,3^{5},10)_9,\ 	(13,3^{8},8)_9,\ 	(13,3^{9},8)_9,\ 	(13,3^{11},5)_9,\ 	(14,3^{5},11)_9,\ 	(14,3^{10},8)_9,\\ 	(16,3^{11},7)_9,\ 	(15,3^{12},8)_9,\ 	(18,3^{11},8)_9,\ 	(19,3^{11},8)_9,\ 	(20,3^{11},9)_9,\ 	(21,3^{11},9)_9.
	\end{align*}
	The constructed trace Hermitian ACD codes with suitable choices of vector $u,$ are listed in Table \ref{Thermitian}.  We compare these codes with the best known linear codes available in Grassl's code table \cite{codetable} and MAGMA \cite{magma} database. Generator matrices for all constructed codes in this article can be found on\\   \small \href{https://drive.google.com/file/d/1sYU1_FM_oeWRVhVmQuSMFJh2-VLhXJcv/view?usp=sharing}{https://drive.google.com/file/d/1sYU1\_FM\_oeWRVhVmQuSMFJh2-VLhXJcv/view?usp=sharing}. 
\end{example}

\begin{table}[h!]
	\begin{center}
%		\caption{ Trace Hermitian ACD codes over the finite field $\mathbb{F}_9=\mathbb{F}_3[w]$ using Theorem \ref{conshermitian}, where $w^2+1=0$. \label{Thermitian}}
		\begin{adjustbox}{width=\textwidth}
			\begin{tabular}{|c|c|c|c|}
				\hline
				$C$&$u$&Constructed code& Best known linear code\\
				\hline
				$(11,3^6,6)$&$(2w+1,w+2,2w+2,w+2,0,0,0,w,0,0,2w)$&$(11,3^7,6)$& $(11,3^7,3)$\\
				%			\hline
				%			$(12,3^7,7)$&$(2w,2w+2,w+1,w+2,0,0,0,0,2w+2,w+2,0,2)$&$(13,3^8,7)^h$& $(13,3^8,4)$\\
				\hline
				$(12,3^{4},10)$&$( 0,0, w^5, w^5, w^5, w^6, 1, 1, w^5,w^3, 2, 1) \ \ (\text{ here } w^2+2w+2=0)$&$(13,3^{5},10)$& $(13,3^5,8)$\\
				\hline
				$(13,3^{7},8)$&$(w+1,w+2,2w+1,2,2w+2,w+2,0,0,0,2w,2w,w+2,2w+2)$&$(13,3^{8},8)$& $(13,3^8,7)$\\
				\hline
				
				$(13,3^{8},8)$&$(w + 1, w + 1, 2, 2w + 1, 2w, 2, w + 2, 0, 0, 0, 2w, 2, 2w + 1)$&$(13,3^{9},8)$& $(13,3^9,4)$\\
				\hline
				$(13,3^{10},5)$&$(2w+2,2w+2,1,w+2,w,0,0,w,w+1,w,2w+1,0,w+2)$&$(13,3^{11},5)$& $(13,3^{11},2)$\\
				\hline
				$(14,3^{4},12)$&$(1, w^2,   2,   w,   2, w^5,   1, w^5,   1,   0,   0,   0, w^3,   1)\ \ \ (\text{ here } w^2+2w+2=0)$&$(14,3^{5},11)$& $(14,3^5,9)$\\
				\hline
				$(14,3^{9},8)$&$(2w + 2, 1, 2, 1, 0,0, 0, 0, 0, 1, 0, 2, 2w + 2, w)$&$(14,3^{10},8)$& $(14,3^{10},7)$\\
				\hline
				$(15,3^{10},7)$&$(1,2w+1,2,0,w,0,0,w,0,2w,w+2,2,1,0,w)$&$(16,3^{11},7)$& $(16,3^{11},4)$\\
				\hline
				$(15,3^{11},8)$&$(w+1,2w +2, w+1, 2w, w, w +2, 0, 0, 0, w, 1, 2w+1, 2, 2w +1, w)$&$(15,3^{12},8)$& $(15,3^{12},7)$\\
				\hline
				$(18,3^{10},8)$&$(2,2w,w,1,1,0,0,0,0,0,0,2,2,0,2w,0,w+2,0)$&$(18,3^{11},8)$& $(18,3^{11},5)$\\
				
				\hline
				$(18,3^{10},8)$&$(w+2,2w+1,w+1,1,1,0,0,0,0,1,w+2,2,2w,0,2w,0,1,0)$&$(19,3^{11},8)$& $(19,3^{11},6)$\\
				\hline
				$(20,3^{10},9)$&$(1,w,2w+2,2w+2,w,0,0,0,0,0,0,0,0,2w,0,0,0,w+1,w,2,w+2)$&$(20,3^{11},9)$& $(20,3^{11},6)$\\
				\hline
				$(20,3^{10},9)$&$(2w+2,w+2,2w+2,2w+2,w,0,0,0,0,0,0,0,0,0,w,0,w,w+1,2,w)$&$(21,3^{11},9)$& $(21,3^{11},7)$\\
				\hline
			\end{tabular}
		\end{adjustbox}
	\end{center}
	\caption{ Trace Hermitian ACD codes over finite field $\mathbb{F}_9=\mathbb{F}_3[w]$ using Theorem \ref{conshermitian}, where $w^2+1=0$. \label{Thermitian}}
\end{table}
\noindent

%%=============================================%%
%% For presentation purpose, we have included  %%
%% \bigskip command. Please ignore this.       %%
%%=============================================%%

\section{Conclusion}\label{conclusion}
Additive codes that trivially intersect with their dual code are known as additive complementary dual (ACD) codes.  In this article, we focused on ACD codes over the finite field $\mathbb{F}_{q^2},$ where $q$ is a prime power. We described ACD codes over $\mathbb{F}_{q^2}$ in terms of generator matrices for the trace Euclidean and the trace Hermitian inner products. We presented a method for constructing ACD codes over $\mathbb{F}_{q^2}$ from linear codes over $\mathbb{F}_q$ and provided some examples to demonstrate these constructions. Furthermore, we devised methods for constructing many ACD codes with parameters $(n,q^{k+1})_{q^2}$ and $(n+1,q^{k+1})_{q^2}$ from a given ACD code with parameters $(n,q^k)_{q^2}$. Using these methods, we constructed several trace Euclidean and  trace Hermitian ACD codes that have a better distance than linear codes over 
$\mathbb{F}_9$ and $\mathbb{F}_4$ of the same size and length.

\bmhead{Acknowledgements}

We would like to express our sincere gratitude to the anonymous reviewers for their insightful comments, and constructive suggestions. Their feedback has been instrumental in improving the quality and clarity of the manuscript. The first author is financially supported by CSIR, New Delhi, Govt. of India, under F. No. 09/086(1407)/2019-EMR-I.

\section*{Statements and Declarations}

%Some journals require declarations to be submitted in a standardised format. Please check the Instructions for Authors of the journal to which you are submitting to see if you need to complete this section. If yes, your manuscript must contain the following sections under the heading `Declarations':

\begin{itemize}
\item Conflict of interest/Competing interests: The authors have no conflicts of interest in this paper. $\bullet$ Data availability: Not Applicable. 
%\item Author contribution: All authors contributed equally to this work.
\end{itemize}

\bibliography{sn-bibliography}% common bib file

\end{document}